\documentclass[useAMS,onecolumn,usenatbib]{mn2e}

\usepackage{graphicx}

\title[Black hole accretion in the Kerr metric as a dynamical system]
{Axisymmetric black hole accretion in the Kerr metric as an autonomous 
dynamical system}
\author[Goswami et al.]
{Sanghamitra Goswami,$^{1}$\thanks{sanghamitra@iitb.ac.in}
Saba Nashreen Khan,$^{2}$\thanks{phs057110@mail2.iitd.ac.in}
Arnab K. Ray$^{3}$\thanks{akr@iucaa.ernet.in}
\newauthor
and Tapas K. Das$^{4}$\thanks{tapas@mri.ernet.in}\\
$^{1}$Department of Physics, Indian Institute of Technology Bombay, 
Powai, Mumbai 400076, India\\
$^{2}$Department of Physics, Indian Institute of Technology Delhi, 
Hauz Khas, New Delhi 110016, India \\
$^{3}$Inter--University Centre for Astronomy and Astrophysics, Post
Bag 4, Ganeshkhind, Pune University Campus, Pune 411007, India\\
$^{4}$Harish--Chandra Research Institute, Chhatnag Road, Jhunsi,
Allahabad 211019, India}

\begin{document}



\maketitle

\label{firstpage}

\begin{abstract}
In a stationary, general relativistic, axisymmetric, inviscid and rotational 
accretion flow, described within the Kerr geometric framework, transonicity 
has been examined by setting up the governing equations of the flow as
a first-order autonomous dynamical system. The consequent linearised
analysis of the critical points of the flow leads to a comprehensive
mathematical prescription for classifying these points, showing that
the only possibilities are saddle points and centre-type points for all
ranges of values of the fixed flow parameters. The spin parameter
of the black hole influences the multitransonic character of the flow, 
as well as some of its specific critical properties. The
special case of a flow in the space-time of a non-rotating black hole,
characterised by the Schwarzschild metric, has also been studied for
comparison and the conclusions are compatible with what has been seen
for the Kerr geometric case.
\end{abstract}


\section{Introduction}
\label{sec1}

From a mathematical perspective, problems in astrophysical accretion
fall under the general class of nonlinear dynamics. This occasions 
no surprise, because accretion, after all, describes the dynamics of 
a compressible astrophysical fluid, and the fundamental governing 
equations of such a problem are nonlinear in nature~\citep{ll87}. 
One of the general issues that is addressed in accretion studies is 
the physics of the accretor itself, whose gravitational attraction 
sustains the global inflow process. This is especially important if 
the accretor is a black hole, which by its very definition is never 
amenable to any direct physical observation, and, therefore, its 
properties can only be known by the gravitational influence it 
exerts on the neighbouring structure of space-time. 

If the accretor is a black hole, the infalling matter has to reach 
the event horizon on a relativistic scale of velocity, and arguments 
in favour of this contention have, by now, gained widespread 
currency~\citep{beg78,bri80,st83,skc90}. Given physically
sensible boundary conditions, this can only imply that  
at one stage the flow of matter will become transonic, i.e. 
its flow speed will grow from subsonic values to supersonic values,
and in doing so, it will at some point match the speed of acoustic
propagation in the fluid. This transition can happen continuously,
as in smooth transonic solutions~\citep{skc90}, or discontinuously, 
as in shocks~\citep{c89,skc90}. 
In accretion studies both possibilities have been subjected to 
concerted investigation. 

Frequently it happens that the transonic feature is exhibited more 
than once in the phase portrait of stationary solutions, i.e. the 
flow will be multitransonic. 
This is particularly true in axisymmetric
rotational flows~\citep{az81,fuk87,c89,skc90,ky94,yk95,par96,
la97,lyyy97,pa97,das02,bdw04,das04,abd06,dbd06}, 
as opposed to steady spherically symmetric flows,
where, as it is well known, physical transonicity occurs only 
once~\citep{beg78,bri80,rb02,mrd07}. 
Physical transonic solutions can be represented mathematically as 
critical solutions in the phase plane of the flow, i.e. they are 
associated with critical points --- alternatively known as fixed 
points or equilibrium points~\citep{js99}. These solutions may even 
pass through 
critical points (as, for instance, a flow through a saddle point). 
In this situation much information about various physical properties
of accretion processes could be gleaned if these critical points are
analysed carefully, which is the central objective of this work. 

In this treatment the nonlinear equations describing the steady, inviscid, 
rotational, axisymmetric flow in the Kerr metric, have been tailored
to form a first-order autonomous dynamical system~\citep{js99}. The 
critical points of the phase trajectories of the flow have been identified
first, following which, a linearised study in the neighbourhood of
these critical points has been carried out. As a consequence of this
exercise, a complete and rigorous mathematical classification scheme
for the nature of the critical points has been derived, and it has
been argued that the critical points can admissibly be only saddle 
points and centre-type points for the kind of conserved, axisymmetric 
and rotational flow under study here. While all of these are principally
the attributes of the hydrodynamical process itself, the influence of 
the black hole (the agent external to the fluid, but driving its flow
nonetheless) has also been
noteworthy to the extent that its intrinsic rotational parameter affects
the character of multitransonicity in general and the properties of an
individual critical point in particular. This is an important result 
to have emerged from the study. 

As a special case of what has been done in the Kerr geometric space-time,
a similar study has also been carried out for the flow in the metric 
of a non-rotating Schwarzschild black hole. The results are entirely 
in keeping with the usual expectations formed on the basis of the 
conclusions drawn from the analysis carried out for the Kerr metric. 

Global understanding of the flow topologies will necessitate a 
full numerical integration of the nonlinear equations of the flow. 
This is never an easy task, but equipped with a clear analytical 
conception of the local features of the critical points of the flow,
it becomes possible to qualitatively anticipate the nature of the 
global behaviour of solutions. In fact, if the system has no more
than two critical points, then with some knowledge of the physical 
boundary conditions of the flow, absolutely correct qualitative 
predictions can be made about the solution topologies~\citep{rb02,rb07}. 

Particularly with respect to accretion on to black holes, this work
follows two earlier works in studying critical properties of accretion
solutions, with the aid of the mathematical tools of a dynamical systems
approach. One work was devoted to discussing the behaviour of critical
points in a steady, axisymmetric, rotational flow, driven towards the
central black hole in the pseudo-Schwarzschild framework~\citep{crd06}. 
The second
work dealt with transonicity and its associated aspects in a fully
general relativistic, spherically symmetric flow described by the
Schwarzschild metric~\citep{mrd07}. The present work on rotational 
flows in the Kerr
metric may be seen as an extension of these earlier efforts, and it
gives an example of the relative mathematical ease with which otherwise
difficult physical questions in general relativistic fluid dynamical 
problems may be confronted. 

\section{The stationary flow in the Kerr metric and its fixed points}
\label{sec2}

Neglecting the self-gravity of 
this general relativistic, axisymmetric, inviscid, stationary, 
compressible hydrodynamic flow in the Kerr geometric space-time, the 
conserved equation for the specific flow energy, $\mathcal E$, 
can be expressed as~\citep{and89,bdw04,dbd06} 
\begin{equation}
\label{eee}
{\mathcal E} = hv_t , 
\end{equation}
which is actually the relativistic analogue of Bernoulli's equation. 
The specific enthalpy, $h$, can be defined as 
\begin{equation}
\label{enthal}
h = \frac{p + \epsilon}{\rho} , 
\end{equation} 
with $\epsilon$, which contains the rest mass density and the internal
energy, being further given by, 
\begin{equation}
\label{epsi}
\epsilon = \rho + \frac{p}{\gamma -1} . 
\end{equation}
The pressure, $p$, is expressed as a function of the density, $\rho$, 
through an equation of state, $p=k \rho^{\gamma}$, from which, under
conditions of constant entropy, $\mathcal S$, the speed of sound is 
defined as 
\begin{equation}
\label{acous}
c_{\mathrm s}^2 =\frac{\partial p}{\partial \epsilon}
\bigg\vert_{\mathcal S} .
\end{equation}
All of these establish a connection between the density, $\rho$, and 
the speed of sound, $c_{\mathrm s}$, as 
\begin{equation}
\label{conden}
\rho = \left[\frac{c_{\mathrm s}^2}{\gamma k
\left(1 - n c_{\mathrm s}^2\right)}\right]^n ,
\end{equation}
with $n$, the polytropic index, being defined as $n=(\gamma -1)^{-1}$. 
This will subsequently lead to an expression of the specific enthalpy 
in terms of $c_{\mathrm s}^2$ as 
\begin{equation}
\label{enthal2}
h = \frac{1}{1-nc_{\mathrm s}^2} . 
\end{equation} 

A further definition gives $v_t$ as 
\begin{equation}
\label{veetee}
v_t = \sqrt{\frac{f(r)}{1-v^2}} , 
\end{equation}
in which $v$ is the radial three-velocity of the corotating 
fluid~\citep{bdw04,dbd06}. The function $f(r)$ is defined as 
\begin{displaymath}
\label{effar}
f(r)=\frac{Ar^2 \Delta}{A^2-4\lambda arA +
\lambda^2 r^2 (4a^2 -r^2 \Delta)} , 
\end{displaymath} 
with $A(r)=r^4 +r^2 a^2 +2ra^2$ and $\Delta (r)=r^2 -2r +a^2$. In all 
of these, the fixed parameters $\lambda$ and $a$ are the sub-Keplerian
specific angular momentum of the flow and the rotating Kerr parameter, 
respectively. Following these definitions, a final expression for the 
relativistic Bernoulli equation will be derived as 
\begin{equation}
\label{bernou}
{\mathcal E}=\frac{1}{1-nc_{\mathrm s}^2}\sqrt{\frac{f(r)}{1-v^2}} . 
\end{equation} 

From the continuity condition, the other governing equation of the flow 
will be obtained as~\citep{bdw04,dbd06} 
\begin{equation}
\label{emdot}
4 \pi \Delta^{1/2} H \rho \sqrt\frac{v^2}{1-v^2} = {\dot m} , 
\end{equation} 
in which the integration constant, ${\dot m}$, is the physical matter 
flow rate. The height of the thin disc flow, $H(r)$, under conditions 
of hydrostatic equilibrium in the vertical direction, 
is expressed as~\citep{bdw04,dbd06} 
\begin{equation}
\label{height} 
H(r)=\sqrt{\frac{2}{\gamma}} r^2 \left[\frac{c_{\mathrm s}^2}
{\left(1 - nc_{\mathrm s}^2 \right) \left\{ \lambda^2 v_t^2
-a^2 \left(v_t-1 \right)\right \}}\right]^{1/2} . 
\end{equation} 

Making use of equations~(\ref{conden}) and~(\ref{height}) in 
equation~(\ref{emdot}), and eliminating the derivatives of $c_{\mathrm s}$
with the help of equation~(\ref{bernou}), it becomes possible, under
the definition that $g_1 (r) = \Delta r^4$ and 
$g_2 (r,v^2) = \lambda^2 v_t^2 - a^2 v_t + a^2$, to arrive
at the relation 
\begin{equation}
\label{dvdr}
\left[ \frac{1}{1-v^2} 
\left( 1 - \frac{\beta^2 c_{\mathrm s}^2}{v^2} \right)
+ \frac{\beta^2 c_{\mathrm s}^2}{g_2} 
\left(\frac{\partial g_2}{\partial v^2}\right) \right]
\frac{\mathrm d}{{\mathrm d}r}(v^2) = \beta^2 c_{\mathrm s}^2
\left[\frac{g^{\prime}_1}{g_1} - \frac{1}{g_2} 
\left(\frac{\partial g_2}{\partial r}\right) \right]
- \frac{f^{\prime}}{f} , 
\end{equation} 
in which $\beta^2 = 2(\gamma +1)^{-1}$, and the primes represent full 
derivatives with respect to $r$. 

From the form of equation~(\ref{dvdr}) it is easy to appreciate that it
is a first-order nonlinear autonomous differential equation, whose 
integration will give the integral solutions in the $r$ --- $v^2$ plane. 
The critical points of these solutions will be derived by the simultaneous
vanishing of the right hand side of equation~(\ref{dvdr}) and the 
coefficient of ${\mathrm d}(v^2)/{{\mathrm d}r}$ in the left hand 
side. This will give the two critical point conditions as 
\begin{equation}
\label{critcon1}
\beta^2 c_{\mathrm{sc}}^2
\left[\frac{g^{\prime}_1(r_{\mathrm c})}{g_1(r_{\mathrm c})} 
- \frac{1}{g_2(r_{\mathrm c},v_{\mathrm c}^2)} 
\left(\frac{\partial g_2}{\partial r}\right)\bigg\vert_{\mathrm c}\right]
- \frac{f^{\prime}(r_{\mathrm c})}{f(r_{\mathrm c})} = 0 
\end{equation}
and 
\begin{equation}
\label{critcon2}
\frac{1}{1-v_{\mathrm c}^2}
\left( 1 - \frac{\beta^2 c_{\mathrm{sc}}^2}{v_{\mathrm c}^2} \right)
+ \frac{\beta^2 c_{\mathrm{sc}}^2}{g_2(r_{\mathrm c},v_{\mathrm c}^2)}
\left(\frac{\partial g_2}{\partial v^2}\right)\bigg\vert_{\mathrm c}=0 , 
\end{equation} 
respectively, with the subscript ``$\mathrm c$" indicating the values at 
the critical points. 

To fix the critical points in terms of the flow parameters, it will be
first necessary to make use of both equations~(\ref{critcon1}) 
and~(\ref{critcon2}) to eliminate $c_{\mathrm{sc}}^2$. Following this, 
some simple algebraic manipulations, with the help of the definition 
of $g_2(r,v^2)$ will make it possible to express $v_{\mathrm c}^2$ 
entirely as a function of $r_{\mathrm c}$, and this will be given by 
\begin{equation}
\label{veecee}
v_{\mathrm c}^2 = \frac{f^{\prime}(r_{\mathrm c}) g_1(r_{\mathrm c})}
{f(r_{\mathrm c}) g_1^{\prime}(r_{\mathrm c})} . 
\end{equation} 
It will then be possible, with the aid of either equation~(\ref{critcon1})
or equation~(\ref{critcon2}), to express $c_{\mathrm{sc}}^2$ as a 
function of $r_{\mathrm c}$ only, and all these conditions, substituted
in equation~(\ref{bernou}), will deliver the roots of $r_{\mathrm c}$ 
in terms of $\mathcal E$, $\lambda$, $\gamma$ and $a$. The critical 
points will, therefore, become fixed in the $r$ --- $v^2$ plane. 

\section{Nature of the fixed points : An autonomous dynamical system}
\label{sec3}

Quite frequently
for any nonlinear physical system, a linearised analytical study of the
properties of the fixed points affords a  
robust platform for carrying out an investigation to understand the 
global behaviour of integral solutions in the phase
portrait. This is especially useful in the absence of any well-prescribed
and general means of solving nonlinear differential equations, which,
perforce, have to be solved numerically. 

It has already been shown that the stationary, axisymmetric, rotational
flow in the Kerr metric can be reduced to a first-order autonomous system,
and as such it lends itself easily to a dynamical systems study of its
fixed points. To do so, it should be necessary to decompose 
equation~(\ref{dvdr}) into two parametrized equations, given by
\begin{eqnarray}
\label{paraeqn}
\frac{\mathrm d}{{\mathrm d}\tau}(v^2) &=& \beta^2 c_{\mathrm{sc}}^2
\left[\frac{g_1^{\prime}}{g_1}-\frac{1}{g_2} 
\left(\frac{\partial g_2}{\partial r}\right)\right]
-\frac{f^{\prime}}{f} \nonumber \\
\frac{{\mathrm d}r}{{\mathrm d}\tau} &=& \frac{1}{1-v^2}
\left(1-\frac{\beta^2 c_{\mathrm{sc}}^2}{v^2}\right)
+\frac{\beta^2 c_{\mathrm{sc}}^2}{g_2}
\left(\frac{\partial g_2}{\partial v^2}\right) ,
\end{eqnarray} 
in which $\tau$ is a mathematical parameter. Since 
equations~(\ref{paraeqn}) are autonomous equations, $\tau$ does not 
explicitly appear in their right hand sides~\citep{js99}. 

This kind of parametrization represents the first step towards 
carrying out a linear stability analysis of the fixed points of 
a nonlinear system, and for the present treatment on disc flows in the
Kerr geometric space-time, this will give a complete classification
scheme for the critical points of the flow. In general fluid dynamics 
problems --- all of which are nonlinear problems --- this approach 
is quite common~\citep{bdp93},
and in the context of accretion studies (which, in its essence, is
the study of a compressible fluid flow), this method has been quite
effectively adopted before~\citep{rb02,ap03,crd06,mrd07}. 
Some earlier works in accretion had also made use of the general 
mathematical aspects of this approach~\citep{mkfo84,mc86,ak89}. 

Making use of the perturbation scheme, $v^2=v_{\mathrm{c}}^2+\delta v^2$, 
$c_{\mathrm{s}}^2=c_{\mathrm{sc}}^2+\delta c_{\mathrm{s}}^2$ and
$r=r_{\mathrm{c}}+\delta r$, along with a modified form of the 
continuity condition, 
\begin{equation}
\label{varsound}
\frac{\delta c_{\mathrm{s}}^2}{c_{\mathrm{sc}}^2} =
{\mathcal A}\,\delta v^2 + {\mathcal B} \,\delta r , 
\end{equation}
in which 
\begin{displaymath}
\label{ae}
{\mathcal A} = -\frac{\gamma-1-c_{\mathrm{sc}}^2}{\gamma +1}
\left[\frac{1}{v_{\mathrm{c}}^2\left(1-v_{\mathrm{c}}^2\right)}
- \frac{1}{g_2(r_{\mathrm{c}},v_{\mathrm{c}}^2)}
\left(\frac{\partial g_2}{\partial v^2}\right)
\bigg \vert_{\mathrm{c}}\right]
\end{displaymath}
and
\begin{displaymath}
\label{bee}
{\mathcal B} = -\frac{\gamma-1-c_{\mathrm{sc}}^2}{\gamma +1}
\left[\frac{g_1^{\prime}(r_{\mathrm{c}})}{g_1(r_{\mathrm{c}})}
- \frac{1}{g_2(r_{\mathrm{c}},v_{\mathrm{c}}^2)}
\left(\frac{\partial g_2}{\partial r}\right)
\bigg \vert_{\mathrm{c}}\right] ,
\end{displaymath}
it is a straightforward exercise to establish a coupled linear 
dynamical system in the perturbed quantities $\delta v^2$ and $\delta r$. 
This is given by 
\begin{eqnarray}
\label{lindyn}
\frac{\mathrm d}{{\mathrm d}\tau}(\delta v^2) &=& \beta^2 c_{\mathrm{sc}}^2
\left[\frac{{\mathcal A}g_1^{\prime}}{g_1} - \frac{\mathcal{AC}}{g_2}
+ \frac{\mathcal{CD}}{g_2^2} - \frac{\Delta_3}{g_2} \right]\, 
\delta v^2 \nonumber \\
& &+ \left[\frac{\beta^2 c_{\mathrm{sc}}^2 g_1^{\prime}}{g_1} 
\left\{{\mathcal B} + \left(\frac{g_1^{\prime \prime}}{g_1^{\prime}} 
- \frac{g_1^{\prime}}{g_1}\right) \right\} - \frac{f^{\prime}}{f}
\left(\frac{f^{\prime\prime}}{f^\prime} - \frac{f^{\prime}}{f}\right)
-\frac{\beta^2 c_{\mathrm{sc}}^2 {\mathcal C}}{g_2} \left({\mathcal B}
- \frac{\mathcal C}{g_2} + \frac{\Delta_4}{\mathcal C} 
\right) \right]\, \delta r \nonumber \\
\frac{{\mathrm d}r}{{\mathrm d}\tau} &=& 
\left[\frac{1}{\left(1-v_{\mathrm{c}}^2\right)^2} 
-\frac{\beta^2 c_{\mathrm{sc}}^2}{v_{\mathrm{c}}^2
\left(1-v_{\mathrm{c}}^2\right)} \left\{ {\mathcal A} 
+ \frac{2v_{\mathrm{c}}^2-1}{\left(1-v_{\mathrm{c}}^2\right)^2}\right\}
+ \frac{\beta^2 c_{\mathrm{sc}}^2 {\mathcal D}}{g_2}\left({\mathcal A}
- \frac{\mathcal D}{g_2} + \frac{\Delta_1}{\mathcal D}\right) 
\right]\, \delta v^2 \nonumber \\
& &+ \left[ -\frac{\beta^2 c_{\mathrm{sc}}^2 {\mathcal B}}
{v_{\mathrm{c}}^2\left(1-v_{\mathrm{c}}^2\right)}
+ \frac{\beta^2 c_{\mathrm{sc}}^2 {\mathcal D}}{g_2} \left({\mathcal B}
-\frac{\mathcal C}{g_2}+\frac{\Delta_2}{\mathcal D}\right)\right]\,\delta r , 
\end{eqnarray}
in which $f$, $g_1$ and $g_2$, all of which are contained in the 
constant coefficients 
of the perturbed quantities, are to be read at the critical points only. 
Explicitly written, all the newly appeared constants in the coefficients 
of equations~(\ref{lindyn}) are to be given as  
\begin{displaymath}
\label{ceeandee}
{\mathcal C} = \left(\frac{\partial g_2}{\partial r}\right)
\bigg \vert_{\mathrm{c}}, \qquad  
{\mathcal D} = \left(\frac{\partial g_2}{\partial v^2}\right)
\bigg \vert_{\mathrm{c}} , 
\end{displaymath}
\begin{displaymath} 
\label{deltas}
\Delta_1 = \frac{\partial}{\partial v^2}
\left(\frac{\partial g_2}{\partial v^2}\right)\bigg \vert_{\mathrm{c}} , 
\qquad
\Delta_2 = \frac{\partial}{\partial r}
\left(\frac{\partial g_2}{\partial v^2}\right)\bigg \vert_{\mathrm{c}} , 
\qquad
\Delta_3 = \frac{\partial}{\partial v^2}
\left(\frac{\partial g_2}{\partial r}\right)\bigg \vert_{\mathrm{c}} ,
\qquad
\Delta_4 = \frac{\partial}{\partial r}
\left(\frac{\partial g_2}{\partial r}\right)\bigg \vert_{\mathrm{c}} .
\end{displaymath} 

Using solutions of the kind $\delta v^2 \sim \exp (\Omega \tau)$ and
$\delta r \sim \exp (\Omega \tau)$, the eigenvalues of the stability 
matrix corresponding to equations~(\ref{lindyn}), can be set down as 
\begin{equation}
\label{eigen}
\Omega^2 = \beta^4 c_{\mathrm{sc}}^4 \chi^2 + \xi_1 \xi_2 , 
\end{equation} 
in which
\begin{displaymath}
\label{chi}
\chi=\left[\frac{g_1^{\prime}{\mathcal A}}{g_1}
-\frac{\mathcal{AC}}{g_2}+\frac{\mathcal{CD}}{g_2^2}
-\frac{\Delta_3}{g_2}\right]
=\left[\frac{\mathcal B}{v_{\mathrm c}^2\left(1-v_{\mathrm c}^2\right)}
-\frac{\mathcal{BD}}{g_2}
+\frac{\mathcal{CD}}{g_2^2}-\frac{\Delta_2}{g_2}\right] , 
\end{displaymath} 
\begin{displaymath}
\label{xi1}
\xi_1=\frac{\beta^2 c_{\mathrm{sc}}^2 g^{\prime}_1}{g_1}
\left[{\mathcal B}+ \frac{g_1^{\prime\prime}}{g_1^{\prime}}
- \frac{g_1^{\prime}}{g_1} \right]
-\frac{f^{\prime}}{f}\left[\frac{f^{\prime\prime}}{f^{\prime}}
-\frac{f^{\prime}}{f}\right] 
- \frac{\beta^2 c_{\mathrm{sc}}^2 \mathcal{C}}{g_2}
\left[{\mathcal B}-\frac{\mathcal C}{g_2}+
\frac{\Delta_4}{\mathcal C}\right]
\end{displaymath} 
and 
\begin{displaymath}
\label{xi2} 
\xi_2=\frac{1}{\left(1-v_{\mathrm c}^2\right)^2}
- \frac{\beta^2c_{\mathrm{sc}}^2}{v_{\mathrm c}^2
\left(1-v_{\mathrm c}^2\right)}\left[{\mathcal A}+
\frac{2v_{\mathrm c}^2-1}{v_{\mathrm c}^2\left(1-v_{\mathrm c}^2\right)}
\right]+\frac{\beta^2 c_{\mathrm{sc}}^2 \mathcal{D}}{g_2}
\left[\mathcal{A}-\frac{\mathcal D}{g_2}+\frac{\Delta_1}{\mathcal D}\right]  
\end{displaymath} 
with $f$, $g_1$ and $g_2$ to be read once again at the critical points. 

The form of equation~(\ref{eigen}) indicates that the critical points
can only be either saddle points (when $\Omega^2 > 0$) or centre-type 
points (when $\Omega^2 < 0$), which is just what they should be for 
a system that is conservative in nature~\citep{js99}. The properties 
of a critical
point could be ascertained by making use of the critical point coordinates, 
$(r_{\mathrm c},v_{\mathrm c}^2)$, in equation~(\ref{eigen}), to find 
the corresponding value of $\Omega^2$, and more especially, its sign. 
This knowledge, along with known and physically meaningful boundary 
conditions of the flow, will 
give a clear idea of the local behaviour of the integral solutions in
the vicinity of the critical points. If the system is simple enough, 
i.e. if it has one or at most two critical points, then a complete
qualitative impression of the global behaviour of the solutions can 
be obtained from this simple analytical exercise~\citep{rb02,rb07}. 
Occasionally these situations can actually arise in thin disc flows 
(relativistic or otherwise) for certain values of the relevant flow 
parameters~\citep{dbd06}. And this is almost certainly true for 
relatively simple spherically symmetric flows~\citep{rb02,mrd07}. 

\section{The Schwarzschild limit}
\label{sec4}

The study carried out so far has been in the Kerr geometric space-time.
Although this is a very general case of a relativistic astrophysical 
flow, of no less interest --- especially from a theoretical viewpoint 
--- is the case of relativistic axisymmetric flows in a spherically 
symmetric metric. This limit --- the Schwarzschild limit --- is to be 
achieved by simply making the Kerr parameter vanish ($a=0$) in 
equations~(\ref{bernou}),~(\ref{emdot}) and~(\ref{height}).
This will make the black hole a non-rotating accretor, and in the 
space-time described by its gravity, the relativistic Bernoulli 
equation will be expressed as 
\begin{equation}
\label{schware}
{\mathcal E} = \frac{1}{1-nc_\mathrm{s}^2}
\sqrt{\frac{r^2\left(r-2\right)}{\left(1-v^2\right)\left[r^3-\lambda^2
\left(r-2\right)\right]}} , 
\end{equation}
while the thickness of the disc, $H(r)$, will go as~\citep{dbd06} 
\begin{equation}
\label{schwarh}
H(r) = \sqrt{\frac{2}{\gamma}}\frac{r^2 c_\mathrm{s}}{\lambda}
\left[\frac{\left(1-v^2\right)
\left\{r^3-\lambda^2\left(r-2\right)\right\}}{\left(1-nc_\mathrm{s}^2\right)
r^2\left(r-2\right)}\right]^{1/2} . 
\end{equation}
The equation of continuity will retain the same form as equation~(\ref{emdot})
even in the Schwarzschild metric, and making use of equations~(\ref{conden})
and~(\ref{schwarh}) in equation~(\ref{emdot}), it should be possible to
recast the equation of continuity in a suitable form as 
\begin{equation}
\label{schwaracous}
\frac{\mathrm d}{{\mathrm d}r}(c_\mathrm{s}^2) = -2 c_\mathrm{s}^2 
\left(\frac{\gamma -1 - c_\mathrm{s}^2}{\gamma +1}\right) 
\left[\frac{1}{2v^2}\frac{\mathrm d}{{\mathrm d}r}(v^2) + f_1(r)\right] , 
\end{equation}
with $f_1$ defined as 
\begin{displaymath}
\label{eff1}
f_1(r) = \frac{3r^3 - 2 \lambda^2 r + 3 \lambda^2}
{r^4 - \lambda^2 r \left(r -2\right)} . 
\end{displaymath} 
Following this, it will become quite easy to 
derive an expression for the gradient of solutions in the $r$ --- $v^2$
plane, and this will read as
\begin{equation}
\label{dvdrsch}
\left[\frac{1}{1-v^2} - \frac{\beta^2 c_\mathrm{s}^2}{v^2} \right]
\frac{\mathrm d}{{\mathrm d}r}(v^2) = 2 \beta^2 c_\mathrm{s}^2
f_1(r) - 2f_2(r) 
\end{equation} 
with $f_2$ being given further by the definition
\begin{displaymath}
\label{eff2}
f_2(r) = \frac{2r - 3}{r\left(r - 2 \right)}
- \frac{2r^3 - \lambda^2 r + \lambda^2}
{r^4 - \lambda^2 r \left(r -2\right)} . 
\end{displaymath}
The critical points (labelled by the subscript ``$\mathrm c$") could be
read from the critical conditions delivered by equation~(\ref{dvdrsch}) as
\begin{equation}
\label{schcrit}
\frac{v_\mathrm{c}^2}{1-v_\mathrm{c}^2} = \beta^2 c_\mathrm{sc}^2
= \frac{f_2(r_\mathrm{c})}{f_1(r_\mathrm{c})} . 
\end{equation}
From the form of the critical points in the foregoing expressions, it is
quite evident that with the help of equation~(\ref{schware}), the critical
point coordinates can be fixed in terms of $\mathcal E$, $\lambda$ and
$\gamma$. To understand the behaviour of the critical points, 
equation~(\ref{dvdrsch}) would have to be parametrized according to 
the prescription outlined in Section~\ref{sec3}. This will entail writing 
\begin{eqnarray}
\label{schpara}
\frac{\mathrm d}{{\mathrm d}\tau}(v^2) &=& 2 \beta^2 c_\mathrm{s}^2
f_1(r) - 2f_2(r)\nonumber \\
\frac{{\mathrm d}r}{{\mathrm d}\tau} &=& \frac{1}{1-v^2} 
- \frac{\beta^2 c_\mathrm{s}^2}{v^2} , 
\end{eqnarray} 
following which, imposing small first-order perturbations about the 
critical point coordinates, the eigenvalues of the stability matrix
of the resulting linearised coupled dynamical system, will be derived as
\begin{equation}
\label{scheigen}
\Omega^2 = \left(f_1 + f_2 \right)^2 \left[\left(2\frac{\gamma-1}{\gamma+1}
- \frac{f_2}{f_1} \right)^2 + \frac{2}{f_1} \left(\frac{2\gamma}{\gamma +1}
+ \frac{1}{2}\frac{f_2}{f_1}\right)\left\{ \frac{f_1^{\prime}}{f_1}
- \frac{f_2^{\prime}}{f_2} - f_1\left(2\frac{\gamma-1}{\gamma+1} -
\frac{f_2}{f_1} \right)\right\}\right] , 
\end{equation}
in which the arguments of the functions $f_1$ and $f_2$ will be 
$r_\mathrm{c}$, with the primes representing full derivatives with 
respect to $r$, as usual. With each physically admissible value 
of $r_\mathrm{c}$,
the nature of the corresponding critical point will be known from 
equation~(\ref{scheigen}), and just like the flow in the Kerr metric,
the critical points will be seen to be only either saddle points or
centre-type points. They could not have been very different, since the
treatment on flows in the Schwarzschild metric is anyway a special case
of the flow in Kerr space-time. Nevertheless, flows in Schwarzschild
geometry merit a separate investigation in their own right, and indeed,
for spherically symmetric flows, this special case affords a very clear
pedagogical model to understand various accretion-related phenomena, 
with some surprisingly new features revealed~\citep{mrd07}. 

\section{Parameter dependence of multitransonicity : General features}
\label{sec5}

\begin{figure}
\begin{center}
\includegraphics[scale=0.3, angle=-90]{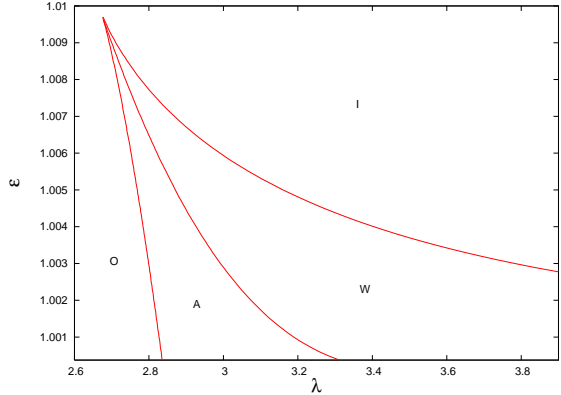}
\caption{\label{f1} \small{Regions of multitransonicity for both
accretion (marked $\mathrm A$) and wind (marked $\mathrm W$) in the 
parameter space of $\mathcal E$ and $\lambda$. The labels $\mathrm I$
and $\mathrm O$ indicate the regions of the lone inner and outer 
critical points, respectively. The plot has been generated for 
$a=0.3$ and $\gamma = 1.33$.}}
\end{center}
\end{figure}

In the two foregoing sections, it has been discussed that for the 
disc flow in the Kerr metric (and for its Schwarzschild limit as 
well), the eigenvalues of the stability matrix associated with each
of the critical points can be fixed in terms of the relevant parameters
of the flow $\mathcal E$, $\lambda$, $\gamma$ and $a$. This obviously
implies that the critical behaviour of the flow can actually be 
determined by these parameters. Multitransonicity is one very important
critical feature of the flow and its dependence on the parameters 
$\mathcal E$ and $\lambda$ in this Kerr geometric system (for fixed
values of $\gamma$ and $a$) has been depicted in Fig.~\ref{f1}. 

The region marked by $\mathrm O$ corresponds to a single outer critical 
point at large length scales. This occurs for low values of $\lambda$
and $\mathcal E$. 
It is not difficult to intuitively grasp the reason for this. Gravity 
drives the accretion process, and this is manifested by the growth of
the velocity field. In a rotating flow angular momentum acts against 
the interest of gravity, and depending on the strength of the presence
of angular momentum, the velocity field will develop accordingly. 
Criticality in the flow is achieved when its velocity matches the 
speed of acoustic propagation in the fluid. The resistance raised 
due to the presence of low angular momentum in the rotating fluid can,
therefore, be overcome even at large distances (where gravity is 
comparatively weak). Besides this, a low value of $\mathcal E$
will imply that the compressible fluid is ``cold" and as such it cannot
offer much of a resistance against gravity with the help of its internal 
thermal effects. 
And so once again gravity wins easily, with criticality developing at 
large length scales. All of these features are manifested in the region 
marked $\mathrm O$ in Fig.~\ref{f1}.

On the other hand, when both $\mathcal E$ and $\lambda$ are high, their 
resistive effects can only be overcome with the rotating fluid having 
to fall deeper within the potential well, and, therefore, criticality 
can be attained only at small length scales, in the vicinity of the event 
horizon of the black hole. This feature has been shown in the region 
(corresponding to higher values of $\lambda$ and $\mathcal E$)
marked $\mathrm I$ in Fig.~\ref{f1}. 
All of these arguments can be appreciated much more easily for rotational
flows in the non-relativistic and Newtonian representation. One might
expect that the qualitative character of the physics in this regime, 
should also carry over smoothly to the general relativistic case. At 
least the parameter space representation in Fig.~\ref{f1} does nothing
to make one believe otherwise, because a similar pattern is also  
exhibited for non-relativistic pseudo-Newtonian flows~\citep{das02}. 

The multitransonic aspect of the flow has been shown by the wedge-shaped 
region in 
Fig.~\ref{f1}. This region has been further subdivided into two regions,
marked by $\mathrm A$ (accretion) and $\mathrm W$ (wind). To gain an
understanding of the physical criterion behind this subdivision, it 
should be necessary first to go to equation~(\ref{emdot}), and then 
substitute $\rho$ in it in terms of $c_\mathrm{s}$, with the help of
equation~(\ref{conden}). This process will lead to the defining of a 
new parameter for the flow, $\dot{\mathcal M} = (\gamma k)^n \dot{m}$. 
This newly defined parameter is physically understood to be the entropy
accretion rate. 

Now multitransonicity in this Kerr geometric flow implies the existence
of three critical points, and under the fundamental physical requirement
of accretion being a process whereby a flow solution should connect 
infinity to the event horizon of the black hole, the three critical 
points should be such that there would be two saddle points flanking 
a centre-type point between themselves. Going back to Fig.~\ref{f1},
the criterion to distinguish region $\mathrm A$ (the accretion region)
will be that the entropy accretion rate, 
$\dot{\mathcal M}_\mathrm{in}$, through the inner saddle point must be
greater than the corresponding flow rate, $\dot{\mathcal M}_\mathrm{out}$,
through the outer saddle point. The exact reversal of this argument, 
i.e. $\dot{\mathcal M}_\mathrm{out} > \dot{\mathcal M}_\mathrm{in}$, 
will define the region $\mathrm W$ (the wind region) in Fig.~\ref{f1}. 

\begin{figure}
\begin{center}
\includegraphics[scale=0.3, angle=-90]{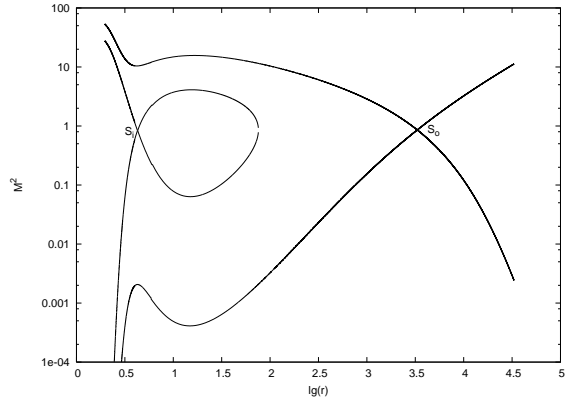}
\caption{\label{f2} \small{Integral solutions of equation~(\ref{dvdr}),
representing region $\mathrm A$ in Fig.~\ref{f1}. 
The vertical axis plots $\mathrm{M}^2$ (since it is convenient to 
measure the bulk velocity against the local speed of sound). Both the
axes have been scaled logarithmically for better graphical resolution 
of the plot. The inner saddle point, $\mathrm{S}_\mathrm{i}$, is 
connected to itself through a homoclinic path. The solutions have 
been generated for $\mathcal E=1.0001$, $\lambda = 3.15$, $a=0.3$ and 
$\gamma = 1.33$.}}
\end{center}
\end{figure}

\begin{figure}
\begin{center}
\includegraphics[scale=0.3, angle=-90]{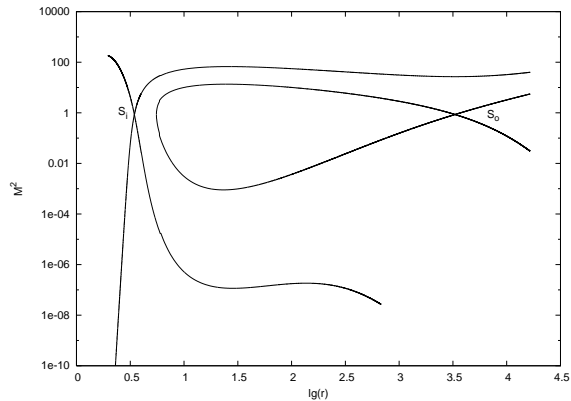}
\caption{\label{f3} \small{Integral solutions of equation~(\ref{dvdr}),
representing region $\mathrm W$ in Fig.~\ref{f1}. The outer saddle 
point, $\mathrm{S}_\mathrm{o}$, is on a homoclinic path. The solutions 
have been generated for $\mathcal E=1.0001$, $\lambda = 3.6$, $a=0.3$ 
and $\gamma = 1.33$.}}
\end{center}
\end{figure}

The former situation can be understood very clearly from Fig.~\ref{f2},
in which the integral solutions of equation~(\ref{dvdr}) have been drawn
under various boundary conditions. The vertical axis plots 
$\mathrm{M}^2$ (with $\mathrm{M}$, the Mach number, being defined as 
$\mathrm{M} = v/c_\mathrm{s}$) against the radial distance along the 
horizontal axis. The behaviour of the solution passing through the 
inner saddle point, $\mathrm{S}_\mathrm{i}$, (whose associated value 
of $\dot{\mathcal M}$ is greater than its corresponding value for 
the outer saddle point, $\mathrm{S}_\mathrm{o}$) is most interesting. 
It connects the point $\mathrm{S}_\mathrm{i}$ to itself through a loop, 
and hence, to invoke the terminology of plane autonomous dynamical systems, 
this solution is actually a homoclinic path~\citep{js99}. 

While Fig.~\ref{f2} depicts the actual behaviour of the phase trajectories
corresponding to the wedge-shaped region marked by $\mathrm A$ in 
Fig.~\ref{f1}, the converse behaviour of these trajectories, corresponding
to the region $\mathrm W$ (wind) has been shown in Fig.~\ref{f3}. Here 
the governing principle is that the entropy accretion rate pertaining to 
the outer saddle point is greater than the one for the inner saddle point
(i.e. $\dot{\mathcal M}_\mathrm{out} > \dot{\mathcal M}_\mathrm{in}$), and
it is evident from the plot that in this situation the solution passing 
through the outer saddle point, $\mathrm{S}_\mathrm{o}$, is a homoclinic 
path. So a general conclusion that can be drawn is that for a physical 
flow such as the one under study here, multitransonicity will imply the 
existence of a homoclinic path (which is a flow solution that connects a 
saddle point to itself). The quantitative physical guideline to identify 
such a solution is to first identify the saddle point through which the 
entropy accretion rate, $\dot{\mathcal M}$, is greater than what it is
for any of the other ones.  

For accretion in particular, a long-standing understanding has been that
the flow has to take place at the maximum possible rate. One could go 
back to a pioneering work in this subject, in which~\citet{bon52} had
conjectured that since there would be nothing to prevent the accretion
process, it might as well take place at the greatest possible rate, 
implying that the flow would be transonic. While this conjecture was 
made on the basis of the spherically symmetric flow, which has a single
critical point (a saddle point), it would still be relevant for 
multitransonic disc flows. Once, after starting under suitable outer 
boundary conditions, an inflow solution has passed through the outer 
saddle point, it will then undergo a transition, so that the entropy 
accretion rate is increased further by the solution passing through the
inner saddle point. This transition will occur through a standing shock,
and this process defines an unambiguous path for the infalling matter
to reach the event horizon of the black hole. Various studies have dealt
with many questions in this regard~\citep{c89,dbd06}, but these details
will be beyond the scope of the present study, which is devoted only
to the critical aspects of the flow. 

A further interesting issue related to Figs.~\ref{f2} and~\ref{f3} is 
that in broad qualitative terms, one is evidently a reversed image of 
the other. While the solutions in Fig.~\ref{f2} are characterised
by $\dot{\mathcal M}_\mathrm{in} > \dot{\mathcal M}_\mathrm{out}$ through
the saddle points, the defining criterion for solutions in Fig.~\ref{f3} 
is $\dot{\mathcal M}_\mathrm{out} > \dot{\mathcal M}_\mathrm{in}$. It will 
then be reasonable to suggest that the accreting system will go through
a state in which 
$\dot{\mathcal M}_\mathrm{in} = \dot{\mathcal M}_\mathrm{out}$. Since 
integral solutions are allowed to intersect only at the critical points 
of a dynamical system~\citep{js99}, this will mean that for the condition 
$\dot{\mathcal M}_\mathrm{in} = \dot{\mathcal M}_\mathrm{out}$, the two
saddle points in the phase portrait will be connected by two heteroclinic
paths only~\citep{js99}. Drawing the phase solutions in this situation, 
however, will entail a tuning of the flow parameters (and the boundary 
conditions) with infinite numerical precision, 
something that should be quite impossible in practice. In Fig.~\ref{f1}
the curve separating region $\mathrm A$ from region $\mathrm W$ depicts
the condition for heteroclinicity. As a matter of fact, this curve, as 
well as the two other curves bounding the multitransonic region in 
Fig.~\ref{f1}, can all be viewed as the loci of various kinds of 
bifurcation points~\citep{js99}. 

\begin{figure}
\begin{center}
\includegraphics[scale=0.3, angle=-90]{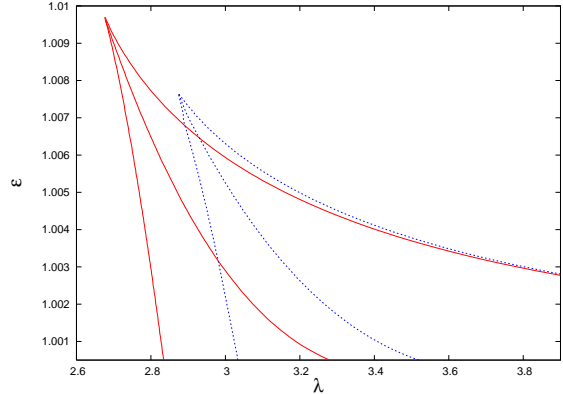}
\caption{\label{f4} \small{A comparison of the multitransonic regions 
in the parameter space of $\mathcal E$ and $\lambda$, for two different
values of $a$, and with $\gamma = 1.33$. The dotted curves (coloured 
blue in the online version) are for $a=0$ (the Schwarzschild limit). 
The continuous curves (coloured red in the online version) are for
$a=0.3$. For the latter case, the shift of the multitransonic region 
towards lower values of $\lambda$ and higher values of $\mathcal E$ is 
quite evident. The area of multitransonicity also increases with the 
increase in the value of $a$.}}
\end{center}
\end{figure}

So far the discussion has dwelt on the critical properties of the flow
for a fixed value of the Kerr rotating parameter, $a$. It should now be
instructive to consider how the variation of $a$ affects the critical 
properties, because, for the case of a rotating black hole, apart from 
its mass (which defines all length scales in the flow), its spin parameter 
will also leave its imprint on the physics of the accretion process. In 
Fig.~\ref{f4}, the way in which the spin parameter influences 
multitransonicity has been shown. The dotted curves delineate the 
region of multitransonicity in the $\lambda$ --- $\mathcal E$ parameter
space for $a=0$ (i.e. the Schwarzschild limit). The continuous curves
indicate the multitransonic region for $a=0.3$. It is very obvious that
the onset of multitransonicity takes place at lower values of $\lambda$
for prograde flows (implied by $a>0$). Besides this, the multitransonic 
region is also 
stretched to higher values of $\mathcal E$. All of these lead to the 
understanding that the Kerr spin parameter (for prograde flows at least)
favourably affects the multitransonic character of the flow. 

\section{Properties of the fixed points in multitransonic flows}
\label{sec6}

\begin{figure}
\begin{center}
\includegraphics[scale=0.3, angle=-90]{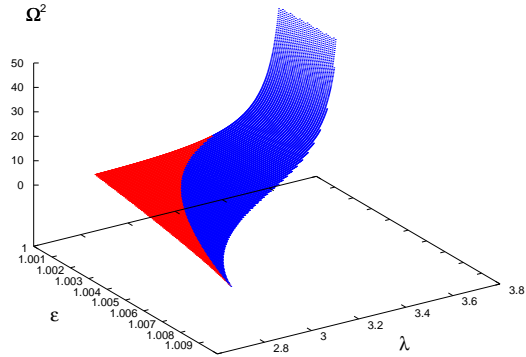}
\caption{\label{f5} \small{All positive values of $\Omega^2$ indicate
that the inner critical point is a saddle point. Its variation
in the parameter space of $\mathcal E$ and $\lambda$ has been shown
for $a=0.3$ and $\gamma = 1.33$. The accretion region is given
by the lightly shaded area (coloured red in the online version), while
the darker region of the surface plot (coloured blue in the online
version) indicates wind.}}
\end{center}
\end{figure}

\begin{figure}
\begin{center}
\includegraphics[scale=0.3, angle=-90]{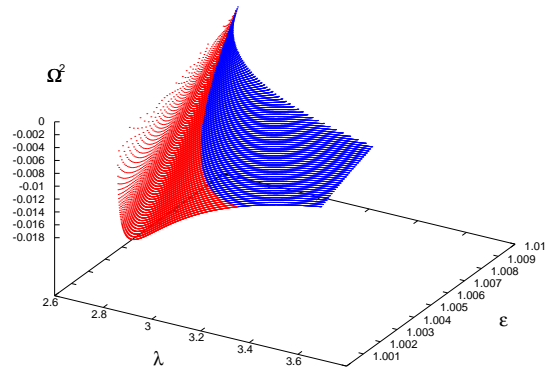}
\caption{\label{f6} \small{The critical point in the middle is a
centre-type point, as the negative values of $\Omega^2$ indicate.
The dependence of $\Omega^2$ on $\mathcal E$ and $\lambda$ has been
shown for $a=0.3$ and $\gamma = 1.33$. The colour scheme
remains the same as before.}}
\end{center}
\end{figure}

\begin{figure}
\begin{center}
\includegraphics[scale=0.3, angle=-90]{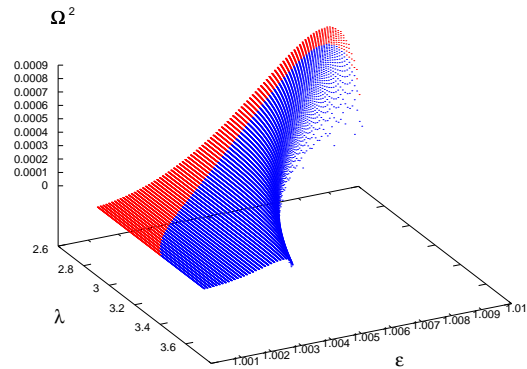}
\caption{\label{f7} \small{The outer critical point is again a saddle
point, since all values of $\Omega^2$ are positive. The dependence of
$\Omega^2$ on the parameters $\mathcal E$ and $\lambda$ has been plotted
for $a=0.3$ and $\gamma = 1.33$.}}
\end{center}
\end{figure}

Various general aspects of multitransonicity have been considered in
detail by now. To derive some quantitative insight about the specific
properties of an individual critical point, however, it will be necessary
to go further. The first thing to do in this regard will be to examine
the behaviour of the eigenvalues of the stability matrix associated with
each critical point. This has to be done by going back to
equation~(\ref{eigen}), which gives a dependence of $\Omega^2$ on the
critical point coordinates. These coordinates, in their turn, have a
dependence on the parameters $\mathcal E$, $\lambda$, $\gamma$ and $a$.
Keeping the last two parameters fixed at $\gamma = 1.33$ and $a=0.3$,
the variation of $\Omega^2$ with respect to $\mathcal E$ and $\lambda$,
has been plotted in Figs.~\ref{f5},~\ref{f6} and~\ref{f7}, for the
inner, the middle and the outer critical points, respectively. In all
these three surface plots the lightly shaded area (coloured red in the
online version) represents accretion, while the dark area (coloured
blue in the online version) represents wind. It will not be difficult
to appreciate that all the three surfaces in
Figs.~\ref{f5},~\ref{f6} and~\ref{f7} will have a two-dimensional
projection on the $\lambda$ --- $\mathcal E$ plot given in Fig.~\ref{f1}.

The sign of $\Omega^2$ indicates the nature of a critical point.
When $\Omega^2$ is positive, it will imply the existence of a saddle
point. And so from Figs.~\ref{f5} and~\ref{f7}, with $\Omega^2$ being
positive all the time in these two plots, it is very much evident that
the inner and the outer critical points are saddle points for a
multitransonic flow. Which is exactly how it should be to make the
whole accretion process feasible, because otherwise there will be no
open path connecting the event horizon of the black hole and the outer
boundary of the flow (which, mathematically speaking, will be at infinity).
Having made a note of this qualitative similarity between these two
critical points, the quantitative differences will also have to be
stressed. The first difference is that the respective values of $\Omega^2$
for either saddle point, differ from those of the other by orders of
magnitude. The inner saddle point behaves more robustly in this regard.
A further difference is that while $\Omega^2$ decreases slightly with
increasing $\mathcal E$ (at a fixed value of $\lambda$) for the inner
saddle point (as Fig.~\ref{f5} shows), the trend is quite the opposite
for the outer saddle point. Here, for a fixed value of $\lambda$, there
is a growth pattern for $\Omega^2$ with increasing $\mathcal E$, as
Fig.~\ref{f7} indicates. This growth is quite noticeable when the value
of $\lambda$ is small. For high values of $\mathcal E$, however, there
is a sharp dip. As opposed to both the critical points in the extremeties,
the critical point in the middle is always a centre-type point, a fact
that is indicated by the negative values of $\Omega^2$, given in
Fig.~\ref{f6}.
Once again, like the outer saddle point, $\Omega^2$ increases with
increasing $\mathcal E$, at a fixed value of $\lambda$.

\begin{figure}
\begin{center}
\includegraphics[scale=0.3, angle=-90]{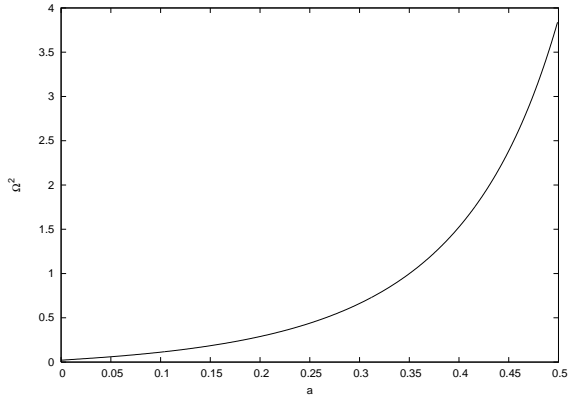}
\caption{\label{f8} \small{For the inner critical point there is 
a monotonic growth pattern for the eigenvalues of the stability matrix, 
$\Omega^2$, with respect to the Kerr parameter, $a$. The other 
relevant flow parameters have been fixed at $\mathcal{E}= 1.00035$,
$\lambda = 3.05$ and $\gamma = 1.33$. Positive values of $\Omega^2$
indicate that the innermost critical point is always a saddle point. 
The growth of $\Omega^2$ with increasing $a$, indicates 
a strengthening of the saddle-like feature of this critical point.}}
\end{center}
\end{figure}

\begin{figure}
\begin{center}
\includegraphics[scale=0.3, angle=-90]{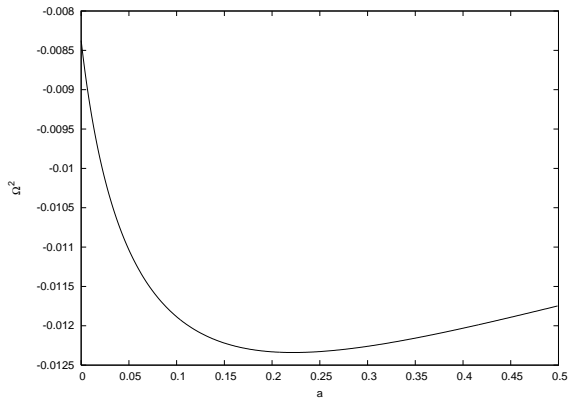}
\caption{\label{f9} \small{Negative values of $\Omega^2$ indicate that
the middle critical point is always a centre-type point. The variation of 
the eigenvalues shows no monotonic behaviour and there is a minimum near 
$a=0.2$. The other flow parameters are $\mathcal{E}= 1.00035$,
$\lambda = 3.05$ and $\gamma = 1.33$.}}
\end{center}
\end{figure}

\begin{figure}
\begin{center}
\includegraphics[scale=0.3, angle=-90]{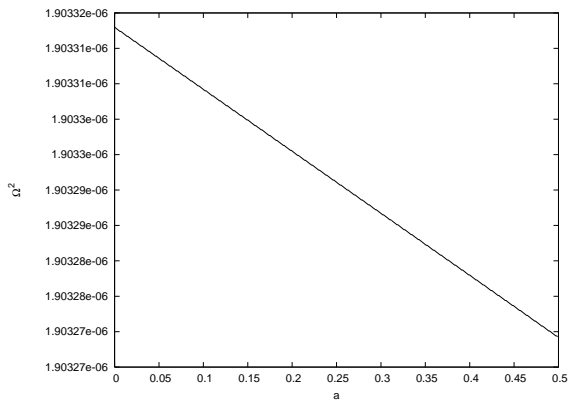}
\caption{\label{f10} \small{The outer critical point is always a 
saddle point, as the positive values of $\Omega^2$ indicate. There is
a steady decay in the magnitude of $\Omega^2$ with increasing $a$, which
implies a gradual weakening of the saddle-like properties. As usual, the
other flow parameters are $\mathcal{E}= 1.00035$, $\lambda = 3.05$ and 
$\gamma = 1.33$.}}
\end{center}
\end{figure}

While Figs.~\ref{f5},~\ref{f6} and~\ref{f7} indicate a dependence of
$\Omega^2$ on $\mathcal E$ and $\lambda$ for fixed values of $a$ and
$\gamma$, it will also be necessary to see how $\Omega^2$ varies with
$a$, having all other parameters fixed. This will reveal how the
mathematical properties of the critical points are affected by an
intrinsic physical property of the black hole. This has been
quantitatively represented in Figs.~\ref{f8},~\ref{f9} and~\ref{f10}.
The two critical points in the extremeties are, of course, saddle
points, but once again, in quantitative terms, they behave in a manner
contrary to each other.
While, for the inner saddle point, $\Omega^2$ grows with increasing
values of $a$ (shown in Fig.~\ref{f8}), there is a steady decline in
the magnitude of $\Omega^2$ for the outer saddle point (shown in
Fig.~\ref{f10}). So, physically speaking, while the presence
of the Kerr parameter augments the properties of the inner saddle point,
at the same time it has an opposite effect on the outer saddle point.
Both these features are, however, manifested in the case of the
centre-type point, for which $\Omega^2$ decreases initially with
increasing $a$, reaches a minimum value (near $a=0.2$), and then
starts to increase. These have all been shown in Fig.~\ref{f9}.

\section{Concluding remarks}
\label{sec7}

While stationary flows are 
interesting enough, studying the dynamic aspects of the accretion 
problem reveals new insight related to transonicity, specifically its 
long-time evolutionary properties and its stability under linearised
time-dependent perturbations. These are relatively less complicated
ventures to undertake for black hole accretion in the pseudo-Schwarzschild
regime, which essentially preserves the simplicity of the Newtonian 
construct of space and time~\citep{crd06}. It has been shown for 
accreting systems, 
both spherically symmetric and axisymmetric, that generating solutions
through a saddle point needs infinitely precise fine-tuning of the 
boundary condition, but the flow easily attains transonicity if its
evolution is traced through time~\citep{rb02,rb07}. While it may be 
proposed that this treatment can be extended to proper general 
relativistic flows, it must also be noted that 
involving explicit time dependence in a general relativistic fluid
flow is always a formidable mathematical problem, more so if the flow
is compressible and rotating. If, on the other hand, this could indeed
be achieved successfully, then a whole host of interesting new features 
would emerge. One issue, related particularly to multitransonic flows
(which arguably will involve more than one saddle point), is the kind 
of solution that the temporal evolution will select, that is to say
which saddle point will the flow finally choose to reach the event horizon
of the black hole, and the physical selection criterion thereof. In 
this connection other questions like variability and chaotic 
behaviour~\citep{dbd06} in the flow might also conceivably be brought 
to the fore. 

\section*{Acknowledgements}

This research has made use of NASA's Astrophysics Data System. 
S. Goswami and S. N. Khan would like to acknowledge the kind 
hospitality provided by HRI, Allahabad, India, under a visiting 
students research programme. The authors are also grateful to   
P. Barai, J. K. Bhattacharjee and S. Nag for much help and some 
useful comments.

\end{document}